# Plasmonic Resonant Optical Nanoswitch

*Andrea Alù and Nader Engheta\**

*Dept. of Electrical and Systems Engineering, University of Pennsylvania, Philadelphia, 19104*

*PA, USA*

**Abstract**

We examine here the anomalous optical response of a nanosphere composed of two conjoined hemispheres of different materials, one of them being plasmonic. At the internal resonance, we show how the light interaction becomes dramatically dependent on the polarization of the local electric field exciting the particle. In particular, its response may be varied from the one of an ideal perfect electric to a perfect magnetic homogeneous sphere through all the possible intermediate stages, by a simple 90° rotation. Applications as optical switch and as a novel nanocircuit element are envisioned.

Light interaction with plasmonic materials gives rise to many anomalous optical phenomena based on localized plasmonic resonance, which have been subject of studies over many years [1]. Their anomalous optical properties have been employed in stained glasses and art masterpieces for centuries [2], but only recently their use has been systematically suggested for technological applications. For instance, the realization of sub-diffractive waveguides [3] and transmission enhancement through sub-wavelength apertures and arrays [4] are based on plasmonic nanoscale resonances, and the future of nanoelectronics is forecasted in the integration of nanoscale plasmonic devices with silicon-based circuits.

One of the most striking plasmonic effects consists of the superlensing properties of a planar plasmonic slab [5]. In fact, a planar slab with permittivity $-\varepsilon_0$, with $\varepsilon_0$ being the permittivity of free-space, may lead to super-resolved images in the near field, due to the excitation of localized surface plasmons [6]. This may also be generalized to a pair of slabs with "complementary" properties [7], i.e., oppositely-signed permittivities and/or permeabilities, whose combined plasmon resonance, supported at their common interface, induces resonant tunneling and super-resolution [7].

Here we analyze a different geometry that may exploit the localized resonance at the internal interface between "complementary" plasmonic and non-plasmonic materials in order to obtain an anomalous optical response, a geometry that consists of the combination of two conjoined hemispheres with oppositely-signed permittivities. Applying a quasi-static (or small radii) mode-matching technique [8], we investigate the anomalous optical properties associated with this combined resonance, which may lead to interesting applications in several areas of optics. At the resonant frequency, in particular, we are able to derive an exact closed-form expression for the field distribution inside and around the nanoparticle, which shows how the resonant response becomes strongly dependent on the orientation of the impinging electric field. As a special case, a hemisphere made of a material with permittivity $-\varepsilon_0$ is analyzed, which may support an anomalous resonance with the surrounding free-space background, in some senses analogous to Pendry's super-resolving lens in the planar geometry [5]. Using a different technique and in a different context, we have predicted some analogous properties for a cylindrical geometry [9], but here we analyze a fully 3D spherical nanoparticle, whose resonant properties may be of interest in different fields of nanoscience and nanotechnology.

Consider the geometry of Fig. 1, i.e., a spherical nanoparticle of radius $a$, smaller than the wavelength of operation $\lambda_0$, composed of two conjoined hemispheres with different permittivities $\varepsilon_{up}$ and $\varepsilon_{down}$, in a free-space background with permittivity $\varepsilon_0$. Without loss of generality, the reference system has been oriented to have the equatorial interface in the $x$-$y$ plane, so that the impinging electric field $\mathbf{E}_0$, forming a generic angle $\gamma$ with respect to the $z$ axis, has its projection on the $x$-$y$ plane oriented along the $x$ axis. This boundary-value problem involves three different dielectrics interfaced with each other and cannot be solved analytically in the more general case. However, it may be solved numerically by considering the mode-matching expansion in terms of spherical harmonics in each region. In the small-radii case, which is of interest here since $a \ll \lambda_0$, this becomes analogous to the numerical solution reported in [8] for a different problem. In the most general case, the electric potential may be written in the different regions as:

$$\phi_{\substack{up\\down}} = E_0 \cos\gamma \sum_{n=0}^{\infty} h_n^{\substack{up\\down}} c_n (r/a)^n P_n(\cos\theta) - E_0 \sin\gamma \sum_{n=0}^{\infty} h_{n+1}^{\substack{up\\down}} d_n (r/a)^n P_n^1(\cos\theta)\cos\varphi$$

$$\phi_0 = E_0 \cos\gamma \sum_{n=0}^{\infty} b_n (r/a)^{-n-1} P_n(\cos\theta) - E_0 \sin\gamma \sum_{n=0}^{\infty} f_n (r/a)^{-n-1} P_n^1(\cos\theta)\cos\varphi +$$

$$-E_0 \cos\gamma (r/a) P_1(\cos\theta) + E_0 \sin\gamma (r/a) P_1^1(\cos\theta)\cos\varphi$$

(1)

where $h_n^{up} = \begin{cases} 1 & n \text{ even} \\ \varepsilon_{down}/\varepsilon_{up} & n \text{ odd} \end{cases}$ and $h_n^{down} = 1$, which results from imposing the boundary conditions on the $x$-$y$ plane, and $(r,\theta,\varphi)$ are spherical coordinates referenced to the nanosphere center. The boundary conditions on the surface of the sphere allow obtaining the unknown sets of coefficients $b_n$, $c_n$, $d_n$ and $f_n$. In particular, using the properties of the Legendre polynomials $P_n$ and $P_n^1$, we can write the following equations for $b_n$, $c_n$:

$$\sum_{n=0}^{\infty} b_n \left[(-1)^{n+l}(1+n+l\varepsilon_{rd}) + \eta_l(1+n+l\varepsilon_{ru})\right] U_{nl} = -U_{1l}\left[(-1)^l(l\varepsilon_{rd}-1) - \eta_l(l\varepsilon_{ru}-1)\right]$$

$$\sum_{n=0}^{\infty} c_n \left[(-1)^{n+l}(1+l+n\varepsilon_{rd}) + \eta_n(1+l+n\varepsilon_{ru})\right] U_{nl} = -U_{1l}\left[1-(-1)^l\right](l+2)$$

(2)

where $\varepsilon_{ru} = \varepsilon_{up}/\varepsilon_0$, $\varepsilon_{rd} = \varepsilon_{down}/\varepsilon_0$ and $U_{nl} = \int_0^1 P_n(x) P_l(x) dx$, which may be evaluated in closed form [8]. An analogous pair of equations may be found for the coefficients $f_n$ and $d_n$, obtainable from Eq. (2) after the substitutions $\eta_t \to \eta_{t+1}$ and $U_{nl} \to U_{nl}^1 = \int_0^1 P_n^1(x) P_l^1(x) dx$, respectively. By truncating the summation in each one of these equations to a given order $N_{max}$ and varying the order $l$ from 0 to $N_{max}$, four systems of equations are obtained, from which it is possible to solve numerically for the unknown coefficients. The convergence of these equations is generally assured, even if near the plasmonic resonances of the sphere (of which the internal one arises for symmetry when $\varepsilon_{up} = -\varepsilon_{down}$) the convergence may become extremely slow and only granted for sufficiently large $n$.

Consider now the special situation in which $\varepsilon_{up} = -\varepsilon_{down}$, i.e., the spherical particle supports the internal plasmonic resonance (notice that this is totally independent of the value of external permittivity $\varepsilon_0$). In this case, the equations for $b_n$ and $f_n$ reduce to:

$$\sum_{n=0}^{\infty} b_n \left[ (1+n+l\varepsilon_{ru}) + (-1)^n (1+n-l\varepsilon_{ru}) \right] U_{nl} = 2l\varepsilon_{ru} U_{1l}, \tag{3}$$

$$\sum_{n=1}^{\infty} f_n \left[ (1+n)(-1+(-1)^n) - l\varepsilon_{ru}(1+(-1)^n) \right] U_{nl} = 2U_{1l}. \tag{4}$$

From Eq. (3), we can easily note that $b_1 = 1$ and $b_n = 0 \; \forall n \neq 1$. Similar considerations apply for Eq. (4), implying $f_1 = -1/2$ and $f_n = 0 \; \forall n \neq 1$. This implies that, at resonance, despite the complexity of the boundary-value problem, a closed-form solution is available for the quasi-static potential in the region outside the sphere:

$$\phi_0 = E_0 \cos\gamma \left[ \left(\frac{a}{r}\right)^2 - \frac{r}{a} \right] \cos\theta - E_0 \sin\gamma \left[ \frac{1}{2}\left(\frac{r}{a}\right)^{-2} + \frac{r}{a} \right] \sin\theta \cos\varphi. \tag{5}$$

We note that the total potential $\phi_0$ does not depend on the specific values of $\varepsilon_{up} = -\varepsilon_{down}$ and it may be simply described by the superposition of two terms, obtained for $\gamma = 0$ ($\mathbf{E}_0$ orthogonal to the interface between the two hemispheres) and for $\gamma = \pi/2$ ($\mathbf{E}_0$ parallel to the interface), respectively, analogous to the cylindrical geometry [10]. For the first term, the potential is identically zero all over the sphere and the potential distribution is exactly equal to the one produced by a perfectly electric homogeneous sphere.

Conversely, the second term corresponds to the potential distribution from a perfectly magnetic homogeneous sphere.

Since we are at the plasmonic resonance between the two hemispheres, it is not surprising that the systems for the internal coefficients $c_n$ and $d_n$ do not properly converge. The reason for this lack of convergence relies on the way in which Eq. (1) is written, which assumes that no electric sources are present inside the nanoparticle. The plasmon resonance, on the other hand, weakens this condition, as it is confirmed by the fact that, when assuming the possibility of having the term with $n = -1$ in the summations for $\phi_{up}$ and $\phi_{down}$, after applying considerations similar to above in solving the corresponding systems of equations, the extra conditions $c_{\pm 1} = \pm 1/\varepsilon_{ru}$, $d_1 = -1$, $d_{-1} = -1/2$, $c_n = d_n = 0 \ \forall |n| \neq 1$ are obtained. This also implies the possibility of closed-form explicit expressions for the potential inside the nanosphere, which may be written as:

$$\phi_{\genfrac{\{}{}{0pt}{}{up}{down}} = E_0 \cos\gamma \left[ (a/r)^2 - r/a \right] \cos\theta / \varepsilon_{\genfrac{\{}{}{0pt}{}{ru}{rd}} - E_0 \sin\gamma \left[ (a/r)^2 /2 + r/a \right] \sin\theta \cos\varphi.$$

(6)

Effectively, the potential inside the nanosphere is produced and sustained by two singular "virtual sources", centered at the origin. The presence of these sub-wavelength singularities, which are effectively the images of the impinging plane wave (more specifically the images of "point sources" at infinity that generate the plane wave excitation), is consistent with our analysis in the cylindrical geometry [9] and is the result of the super-focusing properties of surface plasmons.

Figure 2 shows the electric potential (top) and electric field distribution (bottom) in the $x$-$z$ plane for a nanosphere with $a = 10\,nm$, $\varepsilon_{up} = -16\varepsilon_0$, $\varepsilon_{down} = 16\varepsilon_0$, which may represent, respectively, silver and silicon, around $500 THz$. In this scenario $\gamma = 0$ ($\mathbf{E}_0 \parallel \hat{\mathbf{z}}$) and the surface of the nanosphere is equipotential, with electric field all orthogonal to it, as predicted by Eq. (5). For an external observer, the nanoparticle behaves as a perfectly electric conducting particle, even though no conductive material is employed. Inside the nanoparticle, however, the field is non-zero and a strong circulation of resonant fields is supported by the two singular points (two negative charges for this geometry) at the origin. The dual behavior for the internal field, i.e., the presence of two positive charges at the origin, may be obtained by flipping the particle of $180°$, i.e., by having $\varepsilon_{up} = 16\varepsilon_0$, $\varepsilon_{down} = -16\varepsilon_0$. The field distribution is consistent with the series resonance between a nanoinductor (plasmonic hemisphere) and a nanocapacitor (dielectric hemisphere), as it has been extensively discussed in [9] for the analogous 2D geometry, interpreting this problem in terms of nanocircuit theory [10]-[11]. Consistent with a resonant circuit, an observer outside the particle would simply experience the presence of a perfectly conducting object (short-circuit), without possibly detecting the individual values of $\varepsilon_{up}$ and $\varepsilon_{down}$. However, the electric field inside the pair of hemispheres concentrates towards the singular point at the origin. Although the field is singular at the origin, $\nabla \cdot \mathbf{D} = 0$ everywhere (even at the origin) guaranteeing continuity of displacement current, and charge-free scenario.

By simply rotating the electric field of $90°$, the potential and electric field distribution are dramatically modified, as reported in Fig. 3. Now the equipotential lines are all orthogonal to the surface of the nanosphere and the electric field is tangential to it, exactly as if the nanoparticle were homogeneously filled by a perfect magnetic conductor. Inside the nanoparticle the resonant field circulation is sustained by a virtual singular dipole at the origin, the image of the impinging plane wave focused at the center of the nanoparticle with infinite resolution (in this ideal situation in which losses are neglected). By duality, consistent with the discussion in [9] for the 2D geometry, this situation is completely consistent with the parallel resonance of two hemispheres when interpreted as optical nanocircuit elements.

It is important to note that this resonant spherical particle may indeed act in its entirety as an optical "nanoswitch", which may drastically change its optical response from "short-circuit" (all the impinging displacement current flowing into and out of the sphere) to "open-circuit" (all the impinging displacement current avoiding entering the sphere), by the simple means of a mechanical rotation of 90° (or corresponding rotation of the impinging electric field). This may have important applications in the framework of optical nanocircuits [10]-[11].

It is interesting to note that the same potential and field distributions may be obtained, as a special case, by a simple hemisphere with permittivity $-\varepsilon_0$. In this special situation, the nanoparticle may enter into resonance with the surrounding "complementary" hemisphere in the background, as a sort of hemispherical superlens with its focus at the origin, supporting an analogous effect of drastic dependence of its optical response with the orientation of $\mathbf{E}_0$.

We point out that the total electric field, evaluated as $\mathbf{E} = -\nabla \phi$ from Eq. (5)-(6), satisfies the following relations on the surface of the resonant nanosphere:

$$E_\theta / E_r \big|_{r=a} = \cos \varphi \tan \gamma / 2, \quad E_\varphi / E_r \big|_{r=a} = -\sin \varphi \tan \gamma / (2 \cos \theta). \tag{7}$$

This confirms that the two limiting cases analyzed in Fig. 2 ($\gamma = 0$) and Fig. 3 ($\gamma = 90°$) are characterized by normal and tangential electric fields with respect to the spherical surface, analogous to a perfect electric and magnetic homogeneous sphere, respectively. However, Eq. (7) tells more: it suggests that a rotation of the polarization of electric field (or of the particle) would generate a collection of "intermediate" stages between a perfect electric and a perfect magnetic sphere, for which the angle between the electric field and the normal to the sphere is interestingly proportional to $\tan \gamma$. In some ways, the resonant sphere smoothly changes its optical response as "seen" by the impinging field, from $0$ to $\infty$ as a function of the angle that the external electric field forms with its internal plasmonic interface.

As an example, Fig. 4 reports the cases of a hemisphere with $\varepsilon_{up} = -\varepsilon_0$ (left column) and with $\varepsilon_{down} = -\varepsilon_0$ (right) excited by a field with $\gamma = 45°$. It is interesting to observe how the equipotential lines and the electric field vectors form, in the $x$-$z$ plane, a constant angle with the normal to the spherical surface in both cases (equal to the $\arctan 1/2$, following Eq. (7)), even in the hemispherical region where there is no permittivity contrast with the background. The potential distribution in the region $r > a$ is the same in the two cases, even if, to match the boundary conditions, the fields in the region $r < a$ are reversed from one case to the other. The image singularities at the origin are induced even in the region with $\varepsilon = \varepsilon_0$, associated with the supported plasmon resonance. For an external observer, the whole spherical surface in the $x$-$z$ plane looks homogeneous, and the impinging displacement current is divided into two parts; one flowing into and out of the sphere (related to $E_r$) and the other flowing to the outside (fringing) region (related to $E_\theta$ and $E_\phi$). In particular, the ratio between the current through the sphere and the one in the fringing field is locally determined, on the surface of the sphere, by the ratio between radial and tangential components of the electric field, as described in closed-form in Eq. (7). One notes how this ratio is drastically modified by simply varying the angle $\gamma$, consistent with the nanoswitch functionalities of the resonant plasmonic sphere.

Potential applications of these anomalous properties in tailoring the frequency response of optical nanocircuits may be envisioned. In particular, the possibility of effectively realizing perfect electric or magnetic surfaces, whose response may be even tuned by the applied electric field polarization or by mechanically rotating the nanoparticle (e.g., using optical tweezers), and at frequencies for which metal conductivity is usually low, may be envisioned.

The sensitivity of this resonance to losses and geometry imperfections may be modeled in terms of the optical nanocircuit theory [9]-[11]. In particular, the presence of absorption and imperfections in the involved materials is expected to lower the resonant Q of this system and smooth the singularities predicted by this lossless model. In this context, we have reported in EPAPS [12] the numerical evaluation of (2) for the electric potential distribution consistent with the geometries of Fig. 2 and 3, but considering a permittivity mismatch and including realistic losses, as described in [12]. It can be clearly seen that the concepts described here in the ideal (lossless) resonant limit are qualitatively preserved even when losses and design variations are introduced. The singularities at the origin are now "smoothed" and "spread out" all over the hemispheres' internal interface, due to these imperfections, but the nanoswitch functionalities are still evident. Also our full-wave numerical simulations of this setup considering these aspects show that realistic level of losses in optical materials and current technological limitations may still allow achieving results consistent with the previous discussion.

This work is supported in part by the U.S. Air Force Office of Scientific Research (AFOSR) grant number FA9550-05-1-0442. The authors would like to thank H. Kettunen and A. Sihvola for useful discussions on the mode-matching problem discussed in [8] and M. G. Silveirinha for fruitful discussions.

**REFERENCES**

* To whom correspondence should be addressed. E_mail: engheta@ee.upenn.edu

[1]  M. Kerker, *Applied Optics* **30**, 4699 (1991).

[2]  F. E. Wagner, S. Haslbeck, L. Stievano, S. Calogero, Q. A. Pankhurst and K. -P. Martinek, *Nature* **407**, 691 (2000).

[3]  A. Alù, and N. Engheta, *Phys. Rev. B* **74**, 205436 (2006).

[4]  T. W. Ebbsen, H. J. Lezec, H. F. Ghaemi, T. Thio, and P. A. Wolff, *Nature* **391**, 667 (1998).

[5]  J. B. Pendry, *Phys. Rev. Lett.* **85**, 3966 (2000).

[6]  T. Taubner, D. Korobkin, Y. Urzhumov, G. Shvets, and R. Hillenbrand, *Science* **313**, 1595 (2006).

[7]  A. Alù, and N. Engheta, *IEEE Trans. Antennas Propag.* **51**, 2558 (2003).

[8]  H. Kettunen, H. Wallen, and A. Sihvola, *J. Appl. Phys.* **102**, 044105 (2007).

[9]  A. Alù, A. Salandrino, and N. Engheta, *J. Opt. Soc. Am. B* **24**, 3014 (2007).

[10] N. Engheta, A. Salandrino, A. Alù, *Phys. Rev. Lett.* **95**, 095504 (2005).

[11] N. Engheta, *Science*, **317**, 1698 (2007).

[12] See EPAPS Figure for the electric potential distribution analogous to Fig. 2a ($\gamma = 0$, panel a) and Fig. 2b ($\gamma = \pi/2$, panel b), but with permittivity mismatch and presence of material loss, i.e., $\varepsilon_{up} = \left( -16.1 + i\,0.2 \right)\varepsilon_0$, $\varepsilon_{down} = 16\varepsilon_0$.

**FIGURE LEGENDS**

Figure 1 – (Color online) Geometry of the problem and reference system: two conjoined spherical hemispheres illuminated by a uniform electric field $\mathbf{E}_0$.

Figure 2 – (Color online) Electric potential (top) and electric field distribution (bottom) on the $x$-$z$ plane for the geometry of Fig. 1 with $\varepsilon_{up} = -16\varepsilon_0$, $\varepsilon_{down} = 16\varepsilon_0$, $\gamma = 0$. Darker (more red) colors correspond to larger potential values.

Figure 3 – (Color online) Analogous to Fig. 2, but rotating the electric field to $\gamma = 90°$.

Figure 4 – (Color online) Electric potential (top) and field distribution (bottom) on the $x$-$z$ plane for: (left) $\varepsilon_{up} = -\varepsilon_{down} = \varepsilon_0$, (right) $\varepsilon_{up} = -\varepsilon_{down} = -\varepsilon_0$. In both cases $\gamma = 45°$.

**FIGURES**

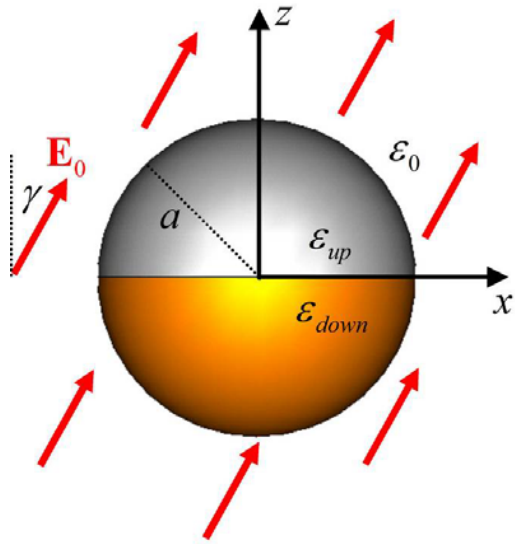

Figure 1

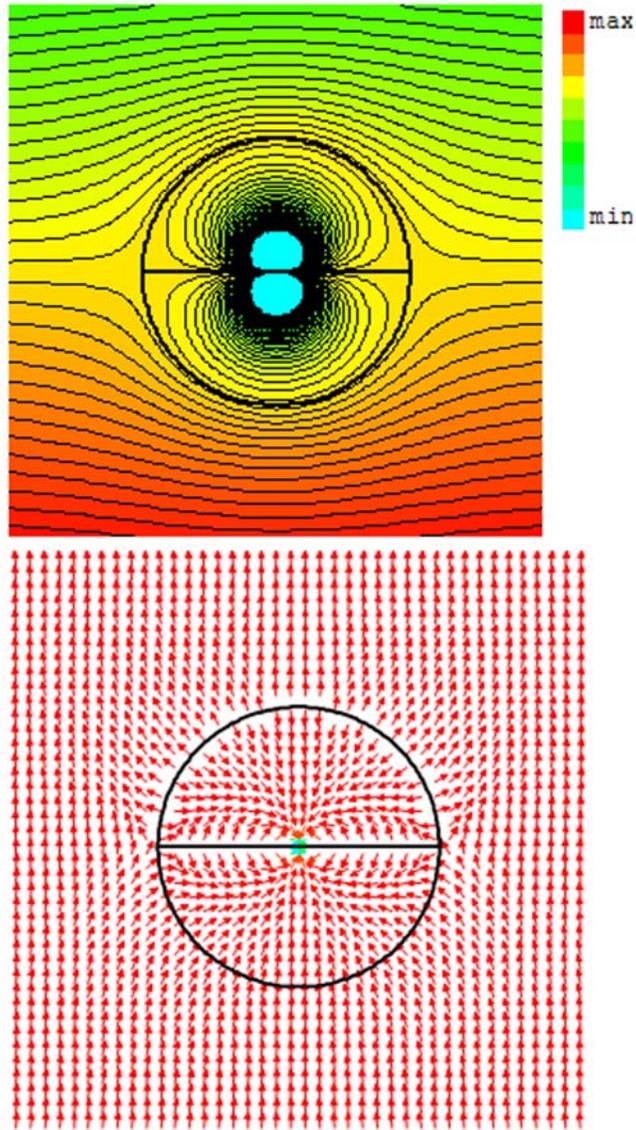

Figure 2

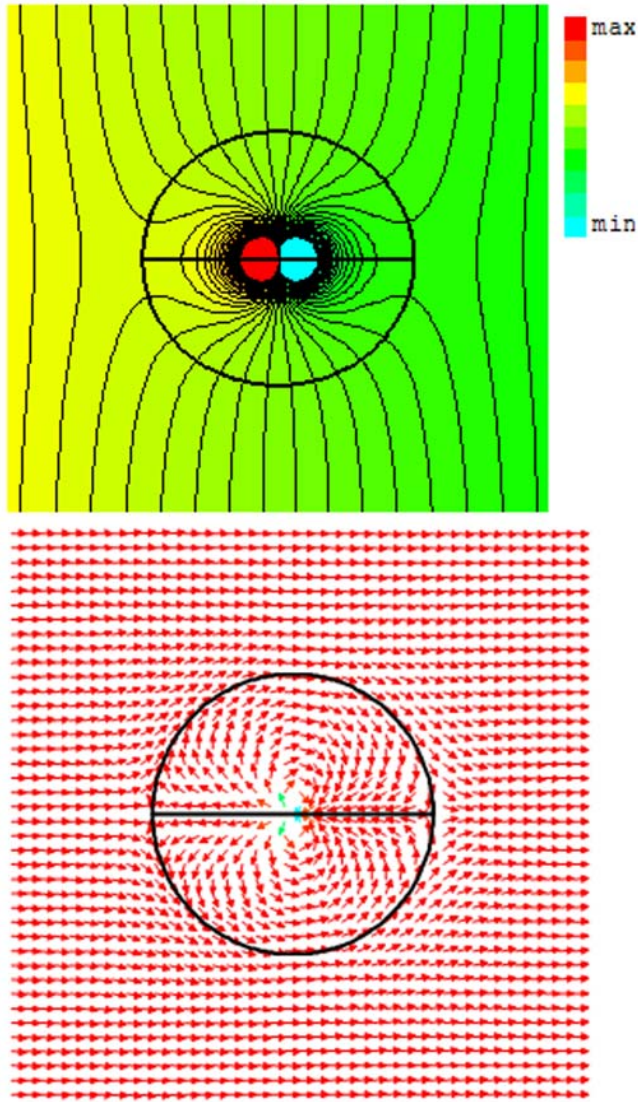

Figure 3

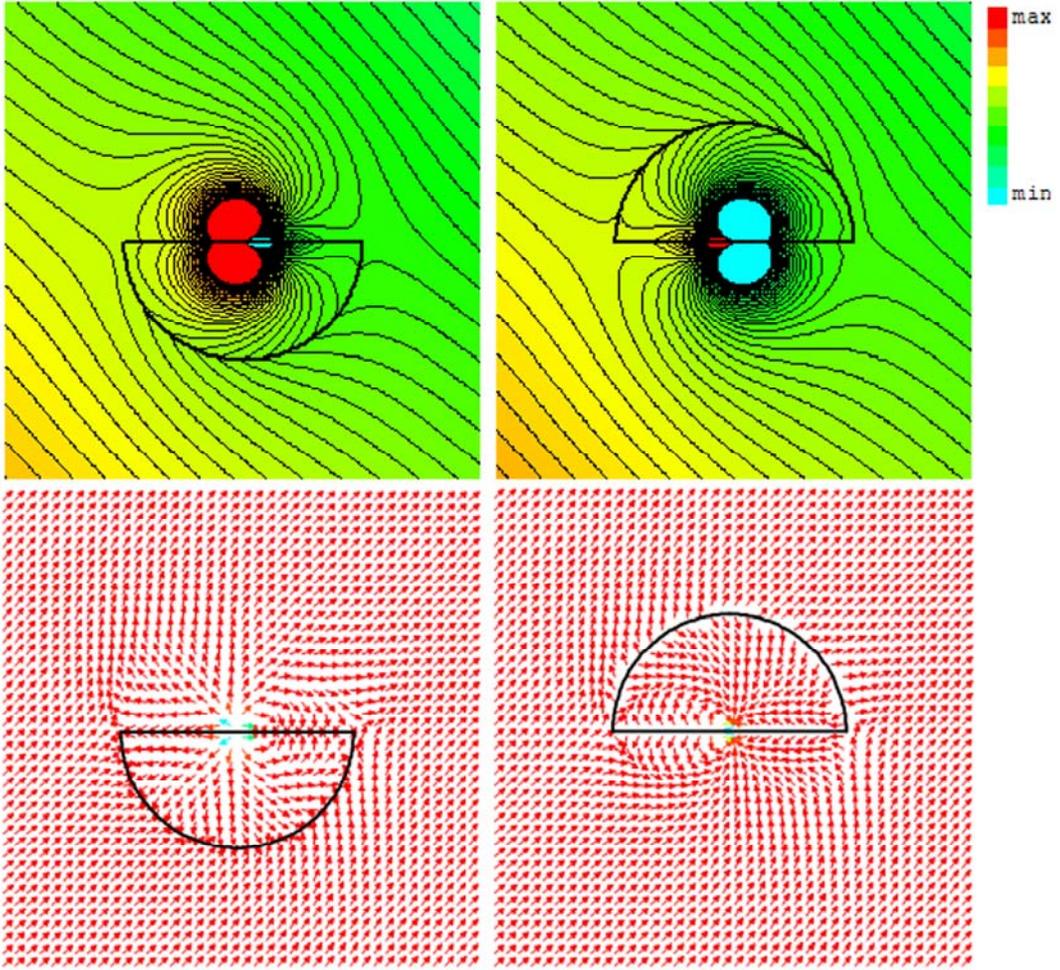

Figure 4

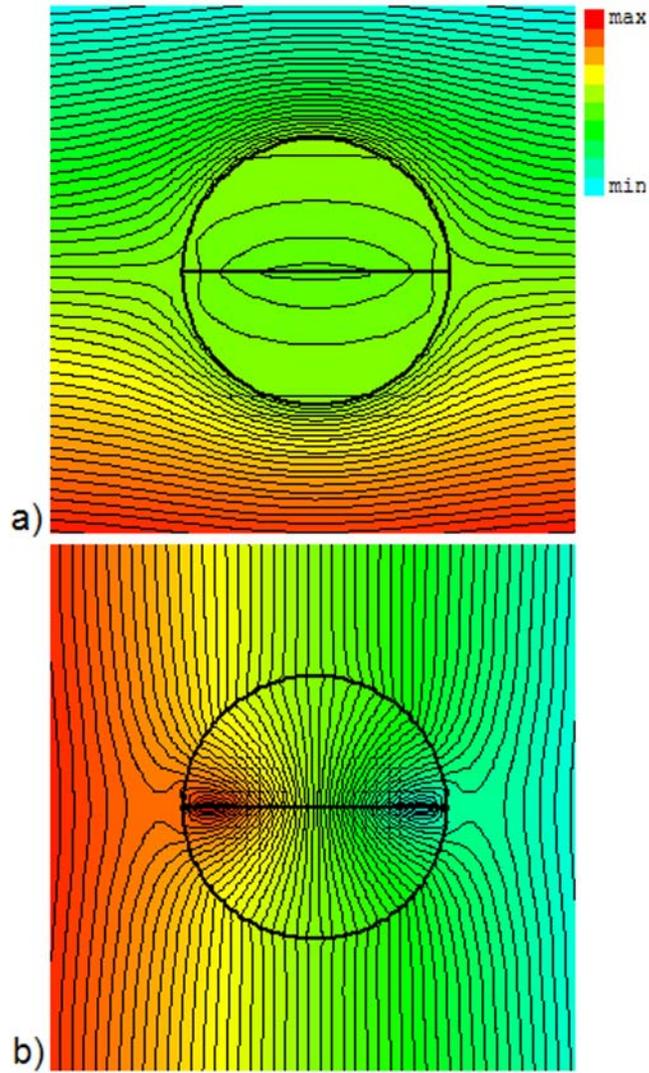

EPAPS Figure [12]